\documentclass[showkeys,showpacs,superscriptaddress]{revtex4-1}
\usepackage{amsmath}
\usepackage{amssymb}
\usepackage{float}
\begin{document}

\title{
Masking singularities in Weyl gravity and Ricci flows
}

\author{
Vladimir Dzhunushaliev
}
\email{v.dzhunushaliev@gmail.com}
\affiliation{
Department of Theoretical and Nuclear Physics,  Al-Farabi Kazakh National University, Almaty 050040, Kazakhstan
}

\affiliation{
Institute of Nuclear Physics, Almaty 050032, Kazakhstan
}

%\affiliation{Institute of Experimental and Theoretical Physics,  Al-Farabi Kazakh National University, Almaty 050040, Kazakhstan}

\affiliation{
Academician J.~Jeenbaev Institute of Physics of the NAS of the Kyrgyz Republic, 265 a, Chui Street, Bishkek 720071, Kyrgyzstan
}

\author{Vladimir Folomeev}
\email{vfolomeev@mail.ru}
\affiliation{
Institute of Nuclear Physics, Almaty 050032, Kazakhstan
}
\affiliation{
Academician J.~Jeenbaev Institute of Physics of the NAS of the Kyrgyz Republic, 265 a, Chui Street, Bishkek 720071, Kyrgyzstan
}
\affiliation{
International Laboratory for Theoretical Cosmology, Tomsk State University of Control Systems and Radioelectronics (TUSUR),
Tomsk 634050, Russia
}

\begin{abstract}
Within vacuum Weyl gravity,  we obtain a solution by which, using different choices of the conformal factor, we derive metrics describing (i)~a bounce of the universe; (ii)~toroidal and spherical wormholes; and (iii)~a change in metric signature. It is demonstrated that singularities occurring in these systems are ``masked''. We give a simple explanation of the possibility of masking the singularities within Weyl gravity. It is shown that in the first and third cases the three-dimensional metrics form Ricci flows. The question of the possible applicability of conformal Weyl gravity as some phenomenological theory in an approximate description of quantum gravity is discussed.
\end{abstract}



\date{\today}

\maketitle

\section{Introduction}

Weyl gravity is a conformally invariant  theory of gravity where classes of conformally equivalent metrics serve as a physical object~\cite{Weyl:gravity}.
For a host of reasons, such theory is not a fundamental theory of gravity, at least on cosmological and solar system scales. Nevertheless, there is a point of view that such theory can be useful in studying various gravitational effects near singularities. For example, in Refs.~\cite{Gurzadyan:2010da,Penrose:1,Penrose:2,Penrose:3}, the idea was proposed according to which classes of conformally equivalent metrics
become a physically important object near a singularity. If this is the case, then such classes may contain both singular and regular metrics.
By the singular metric we mean a metric which makes singular such scalar invariants like the scalar curvature and the squares of the Ricci and Riemann tensors. The existence both of regular and singular metrics within one class of conformally equivalent metrics results in the fact that, despite the divergence
of the aforementioned scalar invariants, the conformally invariant tensors, and hence their squares, will be regular. For instance, we will demonstrate below that, within vacuum Weyl gravity, there exist solutions possessing such properties.
It must be mentioned here that there exists an opposite point of view~\cite{Mannheim:1988dj,OBrien:2011vks},
according to which Weyl gravity theory can be regarded as a viable macroscopic theory of gravitational phenomena. Moreover,  there is a fact that some solutions
of Einsteinian gravity are also exact solutions of Weyl gravity \cite{Li:2015bqa}. Also, in Weyl gravity, there is an interesting approach to explain the nature of dark matter and dark energy~\cite{Flanagan:2006ra}.

In Refs.~\cite{Hooft:2015mzk,THooft:2015jcw}, G.~'t~Hooft considers the idea that ``the conformal symmetry could be as fundamental as Lorentz invariance, and guide us towards a complete understanding of physics at the Planck scale.'' In other words, the conformal invariance may be very important in quantum gravity in describing quantum gravity effects in high curvature regions. This means that Weyl gravity may serve as a phenomenological theory that approximately describes quantum gravity effects in high curvature regions, just as the Ginzburg-Landau theory  is a phenomenological theory of superconductivity.

In Ref.~\cite{Amelino-Camelia:2015dqa}, the idea is proposed that gravity is responsible for breaking the fundamental conformal invariance.
This can be treated as follows: in high curvature regions, the conformal invariance is not violated, but it is violated in going to low enough curvature regions.

In Refs.~\cite{Maldacena:2011mk,Anastasiou:2016jix}, the connection between general relativity and Weyl gravity is under investigation.
It is shown in Ref.~\cite{Maldacena:2011mk} that four-dimensional conformal gravity with a Neumann boundary condition is classically equivalent to ordinary four-dimensional Einstein gravity with a cosmological constant. Ref.~\cite{Anastasiou:2016jix} continues studies in this direction and provides a generic argument on the equivalence between Einstein gravity with a cosmological constant and conformal gravity for Bach-flat spacetimes. In Ref.~\cite{Salvio:2017qkx}, it is shown that agravity can be rewritten as conformal gravity plus two extra scalars with an SO(1,1) symmetry.

In Refs.~\cite{Modesto:2014lga,Modesto:2018def}, the idea is discussed according to which consistent quantum gravity theory must be conformally invariant.
In Ref.~\cite{Modesto:2018def}, on quantum conformal gravity grounds, an approach to resolving the problem of black hole singularity is even suggested.
In Ref.~\cite{narlikar:1977nf}, the idea was pioneered  that singularities arising in general relativity can be eliminated by an appropriate choice of conformal transformation.
The papers~\cite{Bambi:2016wdn,Rachwal:2018gwu} continue studying this subject and suggest an approach to eliminate singularities in Schwarzschild and Kerr solutions.

Concerning Ricci flows, in differential geometry, they are used in studying the topology of differentiable manifolds. Ricci flows define the occurrence of singular points on a manifold; this would lead us to expect that they can be used in gravitational theories when studying such singularities. For example, in Ref.~\cite{Graf:2006mm}, the field equations
are postulated in the form of the Ricci flow equations, and Einstein's theory is included as the limiting case where the flow is absent. In Ref.~\cite{Herrera-Aguilar:2017hzc}, the solutions to the equations for Ricci flows are given, and it is shown that these solutions contain metrics describing a change in metric signature. In Ref.~\cite{Frenkel:2020dic}, Ricci flows are under investigation, and the connection with a path integral in quantum gravity is demonstrated. In Ref.~\cite{Dzhunushaliev:2008cz}, the idea that the occurrence of quantum wormholes in spacetime foam can be described using Ricci flows is discussed.  In Ref.~\cite{Lashkari:2010iy}, Ricci flows are used to study transitions between the AdS and warped AdS vacuum geometries within Topologically Massive Gravity.

Summarizing the above ideas, one can suppose that  Weyl gravity may serve as some phenomenological theory that approximately describes
quantum gravity effects in high curvature regions, and in low curvature regions the conformal invariance is violated. In the present paper,
we would like to demonstrate that, within Weyl gravity, there exists such an interesting feature like masking of some singularities, that in fact is a consequence of quantum gravity, but in the case under consideration this effect is approximately described  by Weyl gravity. By the ``masking of singularities'' we mean the fact that, in Weyl gravity, there can exist the following unusual situation: some tensors (and hence the corresponding scalar invariants) are singular (for example, the Ricci and Riemann tensors),
while at the same time there are tensors which are not singular at the same points. As such tensors, there can be, for example, the Weyl and Bach tensors. From the mathematical point of view, this means that the Weyl tensor is constructed so that the combination of the Riemann and Ricci tensors and of the metric is such that the corresponding singularities eliminate each other. From the physical point of view, this means that Weyl gravity can be treated as an approximate description of quantum gravity effects describing the behavior of spacetime near some singularities.
In this connection it may be also noted that it was shown in Ref.~\cite{Hohmann:2018shl} that
the singularity of a black hole might be removed and replaced by the throat of a wormhole.
It is noteworthy that in $F(R)$ modified gravities, something similar can also exist: it was shown in Ref.~\cite{Bahamonde:2016wmz} that
 in ordinary $F(R)$ gravity a singular cosmology in one frame might be nonsingular in the other frame.

The paper is organized as follows. In Sec.~\ref{Weyl_gravity}, we introduce the Lagrangian and show the corresponding field equations, as well as the  conformally invariant class of metrics which are the solution in Weyl gravity.
For such class of metrics, in Sec.~\ref{bounce}, we discuss a cosmological bounce solution, singularities, and Ricci flows; in Sec.~\ref{signature_change}, we demonstrate the existence of a solution describing a change in metric signature, discuss the corresponding  singularities, and show that the three-dimensional spatial metric is a Ricci flow; in Sec.~\ref{wormoholes}, we obtain toroidal, $T^2$, and spherical, $S^2$, wormholes and study the corresponding  singularities. Finally, in Sec.~\ref{conclus}, we discuss and summarize the results obtained.

\section{Weyl gravity}
\label{Weyl_gravity}

In this section we introduce the Lagrangian and write down the corresponding field equations in Weyl gravity.
The action can be written in the form [hereafter we work in natural units $\hbar = c = 1$ and the metric signature is $(+,-,-,-)$]
\begin{equation}
	\mathcal S = - \alpha_g \int d^4 x \sqrt{-g}
	C_{\alpha \beta \gamma \delta}
	C^{\alpha \beta \gamma \delta} ,
	\label{1_10}
\end{equation}
where $\alpha_g$ is a dimensionless constant,
$
C_{\alpha \beta \gamma \delta} = R_{\alpha \beta \gamma \delta} +
\frac{1}{2} \left(
R_{\alpha \delta} g_{\beta \gamma} -
R_{\alpha \gamma} g_{\beta \delta} +
R_{\beta \gamma} g_{\alpha \delta} -
R_{\gamma \delta} g_{\alpha \beta}\right) +
\frac{1}{6} R \left(
	g_{\alpha \gamma} g_{\beta \delta} -
	g_{\alpha \delta} g_{\beta \gamma}
\right)
$ is the Weyl tensor.

The action \eqref{1_10} and hence the corresponding theory are invariant under the conformal transformations
$$
g_{\mu \nu} \rightarrow f^2(x^\alpha) g_{\mu \nu},
$$
where the function $f(x^\alpha)$ is arbitrary.

The corresponding set of equations in Weyl gravity is
\begin{equation}
	B_{\mu \nu} \equiv 2 C^{\alpha \phantom{\mu \nu} \beta}_
	{\phantom{\alpha} \mu \nu \phantom{\beta} ; \alpha \beta} +
	C^{\alpha \phantom{\mu \nu} \beta}_
	{\phantom{\alpha} \mu \nu \phantom{\beta}} R_{\alpha \beta}
	= 0,
\label{2_10}
\end{equation}
where
$B_{\mu \nu}$ is the Bach tensor.

In what follows we will work with the solutions obtained in Ref.~\cite{Dzhunushaliev:2020dom} and consider in detail how
such solutions can describe the structure of spacetime near singularities. To do this, let us consider the following metric:
\begin{equation}
	ds^2 = f^2 \left(
	t, \chi, \theta, \varphi
	\right) \left\{
	dt^2 - \frac{r^2}{4}
	\left[
	\left( d \chi - \cos\theta d \varphi\right)^2
	+ d \theta^2 + \sin^2 \theta d \varphi^2
	\right]
	\right\} = f^2 \left(
	t, \chi, \theta, \varphi
	\right) \left( dt^2 - r^2 dS^2_3 \right).
	\label{solution_10}
\end{equation}
Here, $dS^2_3$ is the Hopf metric on the unit sphere;
$f \left( t, \chi, \theta, \varphi \right)$ is an arbitrary function; $r$ is a constant;
$0\leq \chi, \varphi \leq 2\pi$ and $0\leq \theta \leq \pi$.
The Bach tensor for such metric is zero; this means that this metric is a solution of Eq.~\eqref{2_10} for Weyl gravity.

In the following sections we will consider some interpretations of this solution and give the analysis of its singular points.

\section{Cosmological bounce solution and inflation }
\label{bounce}

Consider the case with the conformal factor
$$
	f(t, \chi ,\theta ,\varphi ) = f(t) .
$$
Introducing the new time coordinate $d \tau = f(t) dt$,  we have the following expression for the metric~\eqref{solution_10}:
\begin{equation}
  ds^2 = d \tau^2 - \frac{r^2}{4}
	f^2(\tau)
	\left[
		\left( d \chi - \cos\theta d \varphi\right)^2
		+ d \theta^2 + \sin^2 \theta d \varphi^2
	\right] .
\label{3_20}
\end{equation}
If the function $f(\tau)$ is chosen so that $f^2(\tau)$ is an even function and $f(0) = f_0=\text{const}$,
the metric~\eqref{3_20} will describe a universe with a bounce at time $\tau = 0$. If, in addition, one chooses $f(\tau)$ so that asymptotically
$f(\tau) \sim e^\tau$, one will have a cosmological bounce solution with a subsequent inflationary expansion.
This can be done by choosing, for example, $f(t) = 1 / \cos \left(t/t_0 \right)$, as was done in Ref.~\cite{Dzhunushaliev:2020dom}.

Consider the behavior of the metric at the bounce point $t=0$. Since the function $f^2(\tau)$ is assumed to be even, it can be expanded in a Taylor series as
$$
	f^2(t) = f_0 +  t^2 \tilde f(t)
	\equiv f_0 + t^2 \left(
	\frac{f_2}{2} + \frac{f_4}{4!} t^2 + \ldots
	\right),
$$
where $\tilde f(t)$ is an even function. We are interested in the behavior of the metric~\eqref{3_20} at the bounce point $t=0$,
where the scalar invariants have the following expansions:
\begin{align}
&	C_{\alpha \beta \gamma \delta} C^{\alpha \beta \gamma \delta}
	=  0 ,
\quad
	B_{\alpha \beta} B^{\alpha \beta}
	=  0 ,
\label{3_50}\\
&	R	=  -\frac{3 f_2}{f_0} - \frac{6}{r^2 f_0} \propto
	\frac{1}{f_0} \xrightarrow{f_0 \rightarrow 0} \infty,
\quad
	R_{\alpha \beta} R^{\alpha \beta} =
	\frac{3}{4} \frac{f^{(4)}(0)}{f_0^4}
	\xrightarrow{f_0 \rightarrow 0} \infty,
\quad
	R_{\alpha \beta \gamma \delta} R^{\alpha \beta \gamma \delta} =  \frac{3}{2} \frac{f^{(4)}(0)}{f_0^4}
	\xrightarrow{f_0 \rightarrow 0} \infty.
\label{3_80}
\end{align}
If one chooses the function $f(t)$ so that $f_0 = 0$,  an interesting situation takes place at this point: there is a singularity, since the scalar invariants
$
	R, R_{\alpha \beta} R^{\alpha \beta},$ and
	$R_{\alpha \beta \gamma \delta} R^{\alpha \beta \gamma \delta}
$
diverge here. Nevertheless, the Weyl and Bach tensors are constructed so that they are just equal to zero, and hence the corresponding invariants
 $	C_{\alpha \beta \gamma \delta} C^{\alpha \beta \gamma \delta}$ and
		$B_{\alpha \beta} B^{\alpha \beta}$ are also zero.
 \emph{This means that, in Weyl gravity, such singularities are masked!} This is very interesting result, and one might reasonably suppose that it happens due to the fact that Weyl gravity is an approximate description of quantum gravity effects in high curvature regions, i.e., near some singularities.

Thus, the result obtained enables us to say that for small $f_0$ (notice that this parameter corresponds to the size of the universe at the bounce point)
the transition from a contraction stage to expansion is a quantum gravity effect, and it may be approximately described by Weyl gravity.

Next, when the size of the universe increases, the spacetime becomes less curved; correspondingly, quantum gravity effects become negligible and the dynamics of the universe is no longer
adequately described by Weyl gravity; that is, the spacetime becomes classical  and it should be described by general relativity.

There is a very simple explanation why Weyl gravity ``does not see'' the singularity: the reason is that the singularity
arises because of the factor $f^2(\tau)$ in the spatial part of the metric~\eqref{3_20}. Since this factor tends to zero as
$f_0\rightarrow 0$, the volume of the space also goes to zero, thereby the singularities in the scalar invariants
$R, R_{\alpha \beta} R^{\alpha \beta}$, and $R_{\alpha \beta \gamma \delta} R^{\alpha \beta \gamma \delta}$ appear.
But since Weyl gravity is conformally invariant, it ``does not see'' this change in the volume.

Consider now a sequence of spatial metrics from~\eqref{3_20},
\begin{equation}
	dl^2_3 = \frac{r^2}{4}
	f^2(\tau)
	\left[
	\left( d \chi - \cos\theta d \varphi\right)^2
	+ d \theta^2 + \sin^2 \theta d \varphi^2
	\right],
\label{3_90}
\end{equation}
 for $f_0 \rightarrow 0$. This sequence describes the occurrence of the singularity where the invariants \eqref{3_80} go to infinity.
In differential geometry,  this process is described by Ricci flows,
\begin{equation}
	\frac{\partial \gamma_{ij}}{\partial \lambda} = -
	2 R_{ij},
\label{3_100}
\end{equation}
where $\lambda$ is some parameter. The spatial metric tensor $\gamma_{ij}$ from \eqref{3_90} is defined as
\begin{equation}
	\gamma_{ij} = r^2 f^2(\tau, \lambda)
	\tilde{\gamma}_{ij},
\label{3_105}
\end{equation}
where $\tilde{\gamma}_{ij}$ is the metric on the unit three-dimensional sphere in the Hopf coordinates $\chi, \theta, \varphi$;
$R_{ij}$ is the corresponding three-dimensional Ricci tensor with the spatial indices $i, j = 1,2,3$;
the conformal factor $f$ also depends on the parameter $\lambda$.

The Ricci tensor for the metric \eqref{3_90} takes the form
\begin{equation}
	R_{ij} = 2 \tilde{\gamma}_{ij} .
\label{3_110}
\end{equation}
Substituting \eqref{3_105} and \eqref{3_110} in \eqref{3_100}, we get
$$
	\frac{\partial f_0(\lambda)}{\partial \lambda} = -
	\frac{4}{r^2}
$$
with the solution
$$
	f_0(\lambda) = \lambda_0 - \frac{4 \lambda}{r^2} ,
$$
where $\lambda_0$ is an integration constant. This means that the parameter $f_0$, starting from the value $\lambda_0$,
reaches zero value  $f_0 = 0$ when
$
	\lambda = r^2 \lambda_0 / 4
$. For this value, there appears a singularity for the scalar invariants
$
R, R_{\alpha \beta} R^{\alpha \beta},$ and
$R_{\alpha \beta \gamma \delta} R^{\alpha \beta \gamma \delta}
$.

Thus, in this section, we have shown that there are cosmological bounce solutions in Weyl gravity. When the size of the universe decreases, at the bounce point, the scalar invariants
$
	R, R_{\mu \nu} R^{\mu \nu}$, and
	$R_{\mu \nu \rho \sigma}R^{\mu \nu \rho \sigma}
$
diverge but the Weyl invariants
$
	C_{\alpha \beta \gamma \delta} C^{\alpha \beta \gamma \delta}$ and
$B_{\alpha \beta} B^{\alpha \beta}$ remain finite and equal to zero. This enables us to say that, in Weyl gravity, there is a kind of masking of singularities.
It is also shown that the family of such solutions numbered by the size of the universe at the bounce time form the Ricci flow.

The idea that near singularities a conformal  invariance and conformal transformations may be important has been considered in Refs.~\cite{Gurzadyan:2010da,Penrose:1,Penrose:2,Penrose:3}.
The main idea of those papers is that near a singularity the conformal, but not the metric, structure of spacetime is of importance.
And if there exists a conformal factor transferring a metric with a singularity into a metric without a singularity,
from the physical point of view, there is no singularity in such a spacetime. From our point of view, this means that quantum gravity comes into play,
and Weyl gravity is just an approximate description of quantum gravity effects in such a situation.

\section{Passing a singularity with a change in metric signature}
\label{signature_change}

Another interesting example of ignoring a singularity in Weyl gravity is its masking with a change in metric signature. To demonstrate this,  consider the metric
\begin{subequations}
\label{5_10}
\begin{align}
  ds^2 = & d \tau^2 - \frac{r^2}{4} h(\tau)
  \left[
		\left( d \chi - \cos\theta d \varphi\right)^2
		+ d \theta^2 + \sin^2 \theta d \varphi^2
  \right]
\label{5_10_a}\\
  &
  =h(\tau) \left\lbrace
  	\frac{d \tau^2}{h(\tau)} - \frac{r^2}{4}
 	  \left[
 			\left( d \chi - \cos\theta d \varphi\right)^2
 			+ d \theta^2 + \sin^2 \theta d \varphi^2
 	  \right]
  \right\rbrace
\label{5_10_b} \\
  &
  = h(t) \left\lbrace
   	dt^2 - \frac{r^2}{4}
    \left[
			\left( d \chi - \cos\theta d \varphi\right)^2
   			+ d \theta^2 + \sin^2 \theta d \varphi^2
   	  \right]
    \right\rbrace .
\label{5_10_c}
\end{align}
\end{subequations}
In \eqref{5_10_c}, we have introduced $dt = d \tau / \sqrt{h}$ for $\tau > \tau_0$ and
$dt = d \tau / \sqrt{-h}$ for $\tau < \tau_0$. We choose the function $h(\tau)$ so that it changes its sign at some $\tau_0$:
\begin{equation}
	h(\tau) = \begin{cases}
		>0 \text{ by } \tau > \tau_0 , \\
		<0 \text{ by } \tau < \tau_0 .
	\end{cases}
\label{5_20}
\end{equation}
Thus, at $\tau > \tau_0$, the metric signature is Lorentzian, $(+,-,-,-)$, and at $\tau < \tau_0$ it is Euclidean,
 $(+,+,+,+)$. To satisfy the conditions  \eqref{5_20} and simplify calculations, let us choose the function $h(\tau)$ in the form
\begin{equation}
	h(\tau) = \tau \tilde h(\tau) =
	\tau \left(
		h_0 + h_2 \frac{\tau^2}{2!} + \ldots
	\right),
\label{5_30}
\end{equation}
where $\tilde h(\tau)$ is an even function.

At the point $\tau = 0$, we have the following Taylor expansions for the scalar invariants:
\begin{align}
	&C_{\alpha \beta \gamma \delta} C^{\alpha \beta \gamma \delta}
	=  0 ,
\quad
	B_{\alpha \beta} B^{\alpha \beta}
	=  0 ,
\nonumber\\
	& R	\approx  \frac{3}{2 h_0 \tau^3}
	\xrightarrow{\tau \rightarrow 0} \infty ,
\quad
	R_{\alpha \beta} R^{\alpha \beta} \approx
	\frac{9}{4 h_0^2 \tau^6}
	\xrightarrow{\tau \rightarrow 0} \infty ,
\quad
	R_{\alpha \beta \gamma \delta} R^{\alpha \beta \gamma \delta}
	\approx
	\frac{15}{4 h_0^2 \tau^6}
	\xrightarrow{\tau \rightarrow 0} \infty .
\nonumber
\end{align}
Analogous to what was done in the previous section, the Weyl and Bach tensors are nonsingular when passing the point $\tau = 0$,
while the invariants
$
	R, R_{\mu \nu} R^{\mu \nu}$, and
	$R_{\mu \nu \rho \sigma}R^{\mu \nu \rho \sigma}
$ are singular. This means that, as well as in the previous section,  at the point $\tau=0$, where the metric changes its signature, the singularity is masked.

The spatial part of the metric from \eqref{5_10_a} is
$$
	dl^2_3 = \frac{r^2}{4} h(\tau)
	 \left[
		\left( d \chi - \cos\theta d \varphi\right)^2
		+ d \theta^2 + \sin^2 \theta d \varphi^2
	 \right] = r^2 h(\tau)
	 \tilde \gamma_{ij} dx^i dx^j
$$
with the corresponding Ricci tensor
$$
	R_{ij} = 2 \tilde \gamma_{ij}.
$$
In contrast to the cosmological bounce solution considered in Sec.~\ref{bounce} where the quantity $\lambda$ serves as a parameter in the Ricci flow,
here the time coordinate $\tau$ may serve as such a parameter.
%this quantity can be the time coordinate $\tau$ in the case under consideration.
This enables us to write the equation for Ricci flows \eqref{3_100} in the form
$$
	\frac{\partial h(\tau)}{\partial \tau} = -
	\frac{4}{r^2} ,
$$
which gives us the following solution for a Ricci flow:
$$
	h(\tau) = -	\frac{4}{r^2} \tau .
$$
This solution is a special case of the solution to equations for Weyl gravity  \eqref{5_10_a} and \eqref{5_30} for
$\tilde h(\tau)  = \mathrm{const}$.

Thus, in this section, we have shown that, in Weyl gravity, there are solutions describing a change in metric signature.
At the transition point, the scalar invariants
$
	R, R_{\mu \nu} R^{\mu \nu}$, and
	$R_{\mu \nu \rho \sigma}R^{\mu \nu \rho \sigma}
$
go to infinity, but the Weyl invariants
$
	C_{\alpha \beta \gamma \delta} C^{\alpha \beta \gamma \delta}$ and
	$B_{\alpha \beta} B^{\alpha \beta}$
remain finite and equal to zero. As in the case of the solution of Sec.~\ref{bounce},
this effect may be called the masking of singularities in Weyl gravity. Another interesting feature of the solutions with a change in  metric signature obtained here is that
there exists a special solution which is also the Ricci flow for the corresponding spatial part of the metric. Notice that the fact of a change in metric signature in passing through
a singular point has also been pointed out in Ref.~\cite{Herrera-Aguilar:2017hzc}.

\section{Wormholes}
\label{wormoholes}

In the previous sections we have considered a choice of the conformal factor  leading to metrics describing a bounce of the universe and a change in metric signature.
In this section we examine metrics describing wormholes possessing different cross-sections: a $T^2$ torus and a $S^2$ sphere.
In both cases we will obtain the corresponding Ricci flows.

\subsection{Toroidal $T^2$ wormhole}
\label{T2_wormhole}

In this subsection we consider the case where the conformal factor
$f(t, \chi ,\theta ,\varphi )$ depends only on the spatial coordinate $\theta$,
$$
	f(t, \chi ,\theta ,\varphi ) = f(\theta) .
$$
Let us introduce the new spatial coordinate $dx= -f(\theta) d \theta$ so that the function $f(\theta(x))~=~f(x)$
would have a minimum at $x\left( \theta =\pi/2\right) = 0$ and tend to $\pm \infty$ as $x \rightarrow \pm \infty$:
\begin{equation}
	f(0) =  \mathrm{min}, \quad
		f\left( x = \pm \infty\right)  = \pm \infty .
\label{cond_wh}
\end{equation}
In this case we get the following metric:
\begin{equation}
	ds^2 = f^2 \left( \theta (x)\right)
	dt^2 - \frac{r^2}{4} \left\lbrace
	d x^2 + f^2 \left( \theta (x)\right)  \left[
	\left(
		d \chi - \cos \left(\theta(x) \right)  d \varphi
	\right)^2 +
	\sin^2\left( \theta(x)\right)  d \varphi^2
	\right]
	\right\rbrace .
\label{4_20}
\end{equation}
The area of a torus spanned on the coordinates $\chi, \varphi$ is defined by the determinant of the two-dimensional metric
in the square brackets in Eq.~\eqref{4_20}:
$$
	dl^2_2 = f^2 \left( \theta (x)\right)  \left[
		\left(
			d \chi - \cos \left(\theta(x) \right)  d \varphi
		\right)^2 +
		\sin^2\left( \theta(x)\right)  d \varphi^2
		\right] .
$$
Consistent with the conditions \eqref{cond_wh} for the function $f$, it is seen that the area of the torus $S = r^2 f^2(\theta(x))/2$
has a minimum at $\theta = \pi/2$ and goes to infinity as $x \rightarrow \pm \infty$. Also, we require that
$	f(\theta(x)) \sin(\theta(x)) \rightarrow \text{const}$ as $\theta \rightarrow \pi/2$. Then this will mean that we have a toroidal $T^2$ wormhole.

Taking into account that $dx =  - f(\theta) d \theta $ and using the condition
\begin{equation}
	f(\theta) \sin(\theta) =  C = \text{const} ,
\label{4_40}
\end{equation}
we have the following solution for the function $f(x)$, see  Ref.~\cite{Dzhunushaliev:2020dom}:
$$
	f(x) =C \cosh x .
$$
Then the metric \eqref{4_20} takes the form
\begin{equation}
  ds^2 = C^2 \cosh^2 x dt^2 - \frac{r^2}{4} \left\lbrace
    d x^2 + C^2 \cosh^2 x \left[
      \left( d \chi - \tanh x d \varphi \right)^2 +
      \frac{1}{\cosh^2 x} d \varphi^2
    \right]
  \right\rbrace ,
\label{4_50}
\end{equation}
and the coordinate $x$ covers the range $-\infty < x < + \infty$.

Consider now Ricci flows for this case. In Secs.~\ref{bounce} and \ref{signature_change}, we have considered three-dimensional Ricci flows for the spatial parts
of four-dimensional metrics. The argument was that the singularities occurred because of  the fact that the spatial volume vanishes.
In this subsection we consider the case where the area of the wormhole throat  goes to zero;
therefore, we will consider two-dimensional Ricci flows defined on two-dimensional tori which are cross-sections of a wormhole.

Two-dimensional metric for the spacetime metric \eqref{4_50} is
\begin{equation}
	dl^2_2 = - \frac{C^2 r^2}{4} \cosh^2 x \left[
	      \left( d \chi - \tanh x d \varphi \right)^2 +
	      \frac{1}{\cosh^2 x} d \varphi^2
	    \right]= \gamma_{ij} dx^i dx^j ,
	\quad
	x^1 = \chi, x^2 = \varphi,
\label{4_60}
\end{equation}
and Ricci flows should be examined precisely for this two-dimensional metric. In this case a Ricci flow is written as
$$
	\frac{\partial \gamma_{ij}}{\partial \lambda} = -
	2 R_{ij},
$$
where the indices $i, j = \chi, \varphi$ are two-dimensional indices defined on a two-dimensional torus with the metric~\eqref{4_60}. Since the Ricci tensor for the metric \eqref{4_60} is identically zero, $R_{ij} = 0$, this means that the two-dimensional metric~\eqref{4_60} is unchanged in the Ricci flow, as is obvious if we note that  if the condition~\eqref{4_40} is satisfied, we have only one solution~-- the metric~\eqref{4_50}.

One can ignore the condition~\eqref{4_40} and consider wormholes without using it. In that case the even function $f(x)$ has the following Taylor expansion near $x=0$:
$$
	f^2(x) = h(x) = h_0 + x^2 \tilde h(x) =
	h_0 + x^2 \left(
		h_2 + h_4 \frac{x^2}{2!} + \ldots
	\right) .
$$
As $h_0 \rightarrow 0$, there will occur a singularity but, apparently, in this case the factor $f^2(x) \sin^2 x$ before the term with
 $d \varphi^2$ in \eqref{4_20} will also go to zero. This means that we will have a spherical $S^2$ wormhole which will be considered in the next subsection.

Thus, in this subsection, we have considered a toroidal $T^2$ wormhole and shown that for it the Ricci flow is stationary, and thereby singularities are absent in this case.

\subsection{Spherical $S^2$ wormhole}

In the above discussion, we have considered the spacetime with the spatial cross-section in the form of a three-dimensional sphere $S^3$ on which the Hopf coordinates are introduced.
In particular, in the previous subsection, we have shown that, for some special choice of the conformal factor, it is possible to obtain a toroidal $T^2 $ wormhole.
Here, we will demonstrate that, by choosing the standard spherical coordinates on a three-dimensional sphere, it is possible to get a
spherical $S^2 $ wormhole with a cross-section in the form of a two-dimensional sphere $S^2$.

Using the usual spherical coordinates, the spacetime metric can be written in the form
\begin{equation}
	ds^2 = f^2 \left(
		t, \chi, \theta, \varphi
	\right) \left\{
	dt^2 - r^2
	\left[
		d \chi^2	+ \sin^2 \chi \left(
			d \theta^2 + \sin^2 \theta d \varphi^2
		\right)
	\right]
	\right\} = f^2 \left(
		t, \chi, \theta, \varphi
	\right) \left( dt^2 - r^2 dS^2_3 \right),
\label{4_b_10}
\end{equation}
where
$
0 \leqslant \chi \leqslant \pi, 0 \leqslant \theta \leqslant \pi,
0 \leqslant \varphi \leqslant 2 \pi
$ are the angular coordinates on a three-dimensional sphere.

Let us define the conformal factor as
$
	f^2 \left(t, \chi, \theta, \varphi\right) = f^2(\chi)
$.
Then, introducing the new coordinate
$dx =- r f(\chi) d \chi$, we have from~\eqref{4_b_10}:
\begin{equation}
	ds^2 = \left(
		\frac{x^2 + x_0^2}{r x_0}
	\right)^2  dt^2 - d x^2 - \left(
		x^2 + x_0^2
	\right) \left(
			d \theta^2 + \sin^2 \theta d \varphi^2
		\right),
\label{4_b_20}
\end{equation}
where we have used the function
$f(\chi)= x_0 / \left( r \sin^2 \chi \right)$, which gives
	$x = x_0 \cot \chi$ with $- \infty < x < + \infty$. It is evident that this is the metric of a wormhole with the throat radius $x_0$.

For simplicity, we will consider below a $Z_2$ symmetric wormhole. This assumes that after introducing the new coordinate  $x$  [see Eq.~\eqref{4_b_20} above], the function $f(x)$ will be even.
Then the radius of the two-dimensional sphere in the metric~\eqref{4_b_10} can be expanded in a Taylor series in the vicinity of $x=0$
as follows:
\begin{equation}
	f^2(x) \sin^2 \left( \chi(x) \right) =
	h(x) = h_0 + x^2 \tilde h(x) \equiv
	h_0 + x^2 \left(
		h_2 + h_4 \frac{x^2}{2!} + \ldots
	\right) .
\label{4_b_40}
\end{equation}
The parameter $h_0$ defines the area of a two-dimensional sphere at the center of the wormhole  (that is, at the throat). Then the metric~\eqref{4_b_10} takes the form
\begin{equation}
	ds^2 = f^2 (x)
	dt^2 - d x^2	- h(x) \left(
			d \theta^2 + \sin^2 \theta d \varphi^2
		\right).
\label{4_b_50}
\end{equation}

Let us keep track of the behavior of the scalar invariants when a cross-sectional area of the wormhole under consideration goes to zero:
\begin{align}
	& C_{\alpha \beta \gamma \delta} C^{\alpha \beta \gamma \delta}
	=  0 ,
\quad 	B_{\alpha \beta} B^{\alpha \beta}
	=  0 ,
\nonumber\\
&	R	\approx  - \frac{3}{r^2}
	\frac{h_2}{h_0} \xrightarrow{h_0 \rightarrow 0} \infty ,
\quad
	R_{\alpha \beta} R^{\alpha \beta} \approx  \frac{3}{r^4}
	\left(
		\frac{h_2}{h_0^2}
	\right)^2 \xrightarrow{h_0 \rightarrow 0} \infty ,
\quad
	R_{\alpha \beta \gamma \delta} R^{\alpha \beta \gamma \delta}
	\approx  \frac{3}{r^4}
	\left(
		\frac{h_2}{h_0^2}
	\right)^2 \xrightarrow{h_0 \rightarrow 0} \infty .
\nonumber
\end{align}
It is seen from these expressions that the scalar invariants associated with the conformal tensors remain equal to zero, while the scalar invariants
$
	R, R_{\alpha \beta} R^{\alpha \beta}$, and
	$R_{\alpha \beta \gamma \delta} R^{\alpha \beta \gamma \delta}
$ diverge. This means that, in Weyl gravity, when the cross-section of the wormhole decreases, nothing special happens, since the corresponding
invariants do not diverge. From the physical point of view, this process of decrease (or of increase) of the cross-section can be interpreted as an annihilation (or creation) process
of a quantum wormhole in spacetime foam.

As in the case of the toroidal wormhole from Sec.~\ref{T2_wormhole}, here, we will consider two-dimensional Ricci flows, defined now not on two-dimensional tori but
on two-dimensional spheres, which are cross-sections of the wormhole under consideration. The corresponding two-dimensional metric follows from the spacetime
metric~\eqref{4_b_50},
\begin{equation}
	dl_2^2 = - h(x) \left(
		d \theta^2 + \sin^2 \theta d \varphi^2
	\right) = \gamma_{ij} dx^i dx^j =
	h(x) \tilde \gamma_{ij} dx^i dx^j ,
	\quad
	x^1 = \theta, x^2 = \varphi.
\label{4_b_80}
\end{equation}
For it, a Ricci flow is
\begin{equation}
	\frac{\partial \gamma_{ij}}{\partial \lambda} = -
	2 R_{ij},
\label{4_b_90}
\end{equation}
where the indices  $i, j = \theta, \varphi$ are defined on a two-dimensional sphere. The Ricci tensor for the metric~\eqref{4_b_80} is
$$
	R_{ij} = 2 \tilde \gamma_{ij}.
$$
Taking this expression into account and substituting $\gamma_{ij}$ and $\tilde  \gamma_{ij}$ from \eqref{4_b_80} and $h(x)$ from \eqref{4_b_40} in Eq.~\eqref{4_b_90},
we get an equation describing the Ricci flow,
$$
	\frac{\partial h_0}{\partial \lambda} = - 4,
$$
with the solution
$$
	h_0 = \lambda_0 - 4 \lambda .
$$

Thus, in this subsection, we have demonstrated that, in Weyl gravity, there is a family of solutions describing $S^2$ wormholes parameterized by the throat size  $h_0$.
It is shown that when $h_0$ goes to zero, there occur singularities for such invariants like
$
	R, R_{\mu \nu} R^{\mu \nu}$, and
	$R_{\mu \nu \rho \sigma}R^{\mu \nu \rho \sigma}
$. At the same time, the scalar invariants associated with the conformally invariant tensors like
$
	C_{\alpha \beta \gamma \delta} C^{\alpha \beta \gamma \delta}$ and
	$B_{\alpha \beta} B^{\alpha \beta}$ remain regular. This means that, in Weyl gravity, such singularities are masked. It is also shown
that for the $S^2$ wormholes under investigation there are the Ricci flows whose presence can be physically interpreted as the description
of the creation/annihilation process of quantum wormholes in spacetime foam.

\section{Discussion and conclusions}
\label{conclus}

The main purpose of the present paper is to demonstrate that, in Weyl gravity, there is an interesting phenomenon~-- the masking of singularities.
This means that there are solutions for which the scalar invariants
$
R, R_{\alpha \beta} R^{\alpha \beta}$, and
$R_{\alpha \beta \gamma \delta} R^{\alpha \beta \gamma \delta}$ are singular but the tensors employed in Weyl gravity (the Weyl and Bach tensors) remain regular. Perhaps this happens because Weyl gravity can be actually treated as an approximate theory describing quantum gravity effects near singularities, just as the Ginzburg-Landau theory  is a phenomenological theory of superconductivity.
In that case, in the region of strong gravitational fields, quantum gravity is approximately described by Weyl gravity or by some other modified gravity.
By going to a low-curvature region, the conformal invariance violates. Similar ideas concerning the violation of the conformal invariance in quantum Weal gravity
have been considered in Refs.~\cite{Jizba:2020hre,Jizba:2019oaf}.

For better clarity, we would like to emphasize the distinction between the approaches suggested in Refs.~\cite{Modesto:2014lga,Modesto:2018def,Rachwal:2018gwu} and that of stated here.
In the works~\cite{Modesto:2014lga,Modesto:2018def,Rachwal:2018gwu} and other similar papers, it is suggested to quantize some conformally invariant theory of gravitation,
and general relativity follows from it as a classical limit. According to the idea suggested in the present paper, the primary theory is quantized general relativity,
and Weyl gravity arises as an approximate description of some physical system; in the case under consideration, this is a gravitational field near singularities under discussion.
Apparently, consistent quantum gravity will smooth out any singularities and, as it seemed to us, such a process can be approximately described using modified theories of gravity:
Weyl gravity, as in the case considered by us (when the Weyl and Bach tensors vanish), or some other modified theories for black hole singularities
(when the Weyl and Bach tensors are nonzero), for example,  $F(R)$ modified gravities.
Notice also an interesting connection between the solution obtained here within Weyl gravity
and ``the Weyl curvature hypothesis '' proposed in Ref.~\cite{Penrose:3}: in both cases, the Weyl tensor is equal to zero.

An unexpected result of the present study is that we have found the connection between the solutions obtained within Weyl gravity and Ricci flows.
We have shown that for the cosmological bounce solution there is the family of solutions $\gamma(\tau, \lambda)$ indexed by the size of the universe $r^2 f^2(0, \lambda)$ at the bounce time. The element of the family is the metric
$\gamma(\tau, \lambda = \text{const})$ (the spatial part of the four-dimensional metric) which is the solution of the gravitational Weyl equations. In any such family, the metrics $\gamma(\tau = 0, \lambda)$ are a Ricci flow with the Ricci parameter $\lambda$. The solution found in Sec.~\ref{signature_change}, which describes a change in metric signature, is a Ricci flow where the Ricci parameter coincides with the time coordinate $\tau$.

Another interesting result is that all the solutions discussed here refer to one conformally equivalent class of metrics,
where both singular and regular metrics are present.
There, they describe different physical situations:
a bounce of the universe from a singularity with a possible subsequent exponential expansion, toroidal, $T^2$, and spherical, $S^2$, wormholes, and a change in metric signature.

A possible physical explanation of the fact that the metrics under discussion mask the singularities is that Weyl gravity is a phenomenological approximation for microscopic quantum gravity, just as the Ginzburg-Landau theory  is a phenomenological description of superconductivity.

Thus, summarizing the results obtained:
\begin{itemize}
\item Within Weyl gravity, there are obtained four types of solutions which are conformally equivalent each other but describe different physical situations.
\item It is shown that for all these solutions the singularities are masked in the sense that, even though such scalar invariants like the scalar curvature
and the squares of the Ricci and Riemann  tensors are singular, the squares of the Weyl and Bach tensors (which are employed in Weyl gravity) remain regular.
\item It is shown that for these solutions the three/two-dimensional spatial metrics are simultaneously Ricci flows.
\item A possible interpretation of Weyl gravity as a phenomenological theory which approximately describes quantum gravity effects is discussed.
\end{itemize}

\section*{Acknowledgments}
We gratefully acknowledge support provided  by the Program No.~BR10965191 of the Ministry of Education and Science of the Republic of Kazakhstan. We are also grateful to the Research Group Linkage Programme of the Alexander von Humboldt Foundation for the support of this research.


\begin{thebibliography}{99}

\bibitem{Weyl:gravity}
H. Weyl,
``Gravitation und Elekrizit\"at'',
Sitzungsberichte der K\"oniglich Preusischen Akademie der Wissenschaften zu Berlin, 1918, pp. 465-480;
English translation, ``Gravitation and Electricity,'' pp. 24-37 in O'Raifeartaigh's book.

%\cite{Gurzadyan:2010da}
\bibitem{Gurzadyan:2010da}
V.~G.~Gurzadyan and R.~Penrose, Sir,
``Concentric circles in WMAP data may provide evidence of violent pre-Big-Bang activity,''
arXiv:1011.3706 [astro-ph.CO].
%47 citations counted in INSPIRE as of 04 Feb 2021

\bibitem{Penrose:1}
R. Penrose,
{\it  Causality, quantum theory and cosmology}
(In On Space and Time ed. Shahn Majid, Cambridge University Press, Cambridge, 2008) pp. 141-195. % (ISBN 978-0-521- 88926-1)

\bibitem{Penrose:2}
R. Penrose, {\it  The Basic Ideas of Conformal Cyclic Cosmology}
(In Death And Anti-Death, Volume 6: Thirty Years After Kurt G\"odel (1906-1978), Chapter 7, pp. 223-242. Ed. Charles Tandy, Ria University Press, Stanford, Palo Alto, Calif., 2009).  % ISBN 978-1-934297-03-2.

\bibitem{Penrose:3}
R. Penrose,
{\it  Cycles of Time: An Extraordinary New View of the Universe}
(Bodley Head, London, 2010). %(ISBN 9780224080361).

%\cite{Mannheim:1988dj}
\bibitem{Mannheim:1988dj}
P.~D.~Mannheim and D.~Kazanas,
  ``Exact Vacuum Solution to Conformal Weyl Gravity and Galactic Rotation Curves,''
  Astrophys.\ J.\  {\bf 342}, 635 (1989).

%\cite{OBrien:2011vks}
\bibitem{OBrien:2011vks}
J.~G.~O'Brien and P.~D.~Mannheim,
  ``Fitting dwarf galaxy rotation curves with conformal gravity,''
  Mon.\ Not.\ Roy.\ Astron.\ Soc.\  {\bf 421}, 1273 (2012).

%\cite{Li:2015bqa}
\bibitem{Li:2015bqa}
 Y.~D.~Li, L.~Modesto and L.~Rachwa\l{},
  ``Exact solutions and spacetime singularities in nonlocal gravity,''
  JHEP {\bf 1512}, 173 (2015).

%\cite{Flanagan:2006ra}
\bibitem{Flanagan:2006ra}
E.~E.~Flanagan,
  ``Fourth order Weyl gravity,''
  Phys.\ Rev.\ D {\bf 74}, 023002 (2006).

%\cite{Hooft:2015mzk}
\bibitem{Hooft:2015mzk}
 G.~'t Hooft,
  ``Spontaneous breakdown of local conformal invariance in quantum gravity,''
  Les Houches Lect.\ Notes {\bf 97}, 209 (2015).

%\cite{THooft:2015jcw}
\bibitem{THooft:2015jcw}
G.~'t Hooft,
  ``Local conformal symmetry: The missing symmetry component for space and time,''
  Int.\ J.\ Mod.\ Phys.\ D {\bf 24}, no. 12, 1543001 (2015).

%\cite{Amelino-Camelia:2015dqa}
\bibitem{Amelino-Camelia:2015dqa}
 G.~Amelino-Camelia, M.~Arzano, G.~Gubitosi, and J.~Magueijo,
  ``Gravity as the breakdown of conformal invariance,''
  Int.\ J.\ Mod.\ Phys.\ D {\bf 24}, no. 12, 1543002 (2015).

%\cite{Maldacena:2011mk}
\bibitem{Maldacena:2011mk}
J.~Maldacena,
``Einstein Gravity from Conformal Gravity,''
[arXiv:1105.5632 [hep-th]].
%257 citations counted in INSPIRE as of 26 Feb 2021

%\cite{Anastasiou:2016jix}
\bibitem{Anastasiou:2016jix}
G.~Anastasiou and R.~Olea,
``From conformal to Einstein Gravity,''
Phys.\ Rev.\ D {\bf 94}, no. 8, 086008 (2016).

%\cite{Salvio:2017qkx}
\bibitem{Salvio:2017qkx}
A.~Salvio and A.~Strumia,
``Agravity up to infinite energy,''
Eur.\ Phys.\ J.\ C {\bf 78}, no. 2, 124 (2018).

%\cite{Modesto:2014lga}
\bibitem{Modesto:2014lga}
L.~Modesto and L.~Rachwa\l{},
  ``Super-renormalizable and finite gravitational theories,''
  Nucl.\ Phys.\ B {\bf 889}, 228 (2014).

%\cite{Modesto:2016max}
%\bibitem{Modesto:2016max}
%L.~Modesto and L.~Rachwal,
%``Finite Conformal Quantum Gravity and Nonsingular Spacetimes,''
%[arXiv:1605.04173 [hep-th]].
%42 citations counted in INSPIRE as of 08 Apr 2021

%\cite{Modesto:2018def}
\bibitem{Modesto:2018def}
 L.~Modesto and L.~Rachwa\l{},
  ``Finite conformal quantum gravity and spacetime singularities,''
  J.\ Phys.\ Conf.\ Ser.\  {\bf 942}, no. 1, 012015 (2017).

%\cite{narlikar:1977nf}
\bibitem{narlikar:1977nf}
J.~V.~Narlikar and A.~K.~Kembhavi,
  ``Space-Time Singularities and Conformal Gravity,''
  Lett.\ Nuovo Cim.\  {\bf 19}, 517 (1977).

%\cite{Bambi:2016wdn}
\bibitem{Bambi:2016wdn}
C.~Bambi, L.~Modesto, and L.~Rachwa\l{},
  ``Spacetime completeness of non-singular black holes in conformal gravity,''
  JCAP {\bf 1705}, 003 (2017).

%\cite{Rachwal:2018gwu}
\bibitem{Rachwal:2018gwu}
L.~Rachwa\l{},
``Conformal Symmetry in Field Theory and in Quantum Gravity,''
Universe {\bf 4}, no. 11, 125 (2018).
%10 citations counted in INSPIRE as of 08 Apr 2021

\bibitem{Graf:2006mm}
W.~Graf,
``Ricci flow gravity,''
PMC Phys.\ A {\bf 1}, 3 (2007).

\bibitem{Herrera-Aguilar:2017hzc}
R.~Cartas-Fuentevilla, A.~Herrera-Aguilar, and J.~A.~Olvera-Santamaria,
``Evolution and metric signature change of maximally symmetric spaces under the Ricci flow,''
Eur.\ Phys.\ J.\ Plus {\bf 133}, no. 6, 235 (2018).

%\cite{Frenkel:2020dic}
\bibitem{Frenkel:2020dic}
 A.~Frenkel, P.~Horava, and S.~Randall,
  ``Perelman's Ricci Flow in Topological Quantum Gravity,''
  arXiv:2011.11914 [hep-th].
%1 citations counted in INSPIRE as of 12 Feb 2021

%\cite{Dzhunushaliev:2008cz}
\bibitem{Dzhunushaliev:2008cz}
V.~Dzhunushaliev,
``Quantum wormhole as a Ricci flow,''
Int.\ J.\ Geom.\ Meth.\ Mod.\ Phys.\  {\bf 6}, 1033 (2009).

\bibitem{Lashkari:2010iy}
N.~Lashkari and A.~Maloney,
``Topologically Massive Gravity and Ricci-Cotton Flow,''
Class.\ Quant.\ Grav.\  {\bf 28}, 105007 (2011).

%\cite{Hohmann:2018shl}
\bibitem{Hohmann:2018shl}
 M.~Hohmann, C.~Pfeifer, M.~Raidal, and H.~Veerm\"{a}e,
  ``Wormholes in conformal gravity,''
  JCAP {\bf 1810}, 003 (2018).

%\cite{Bahamonde:2016wmz}
\bibitem{Bahamonde:2016wmz}
S.~Bahamonde, S.~D.~Odintsov, V.~K.~Oikonomou, and M.~Wright,
  ``Correspondence of $F(R)$ Gravity Singularities in Jordan and Einstein Frames,''
  Annals Phys.\  {\bf 373}, 96 (2016).

%\cite{Dzhunushaliev:2020dom}
\bibitem{Dzhunushaliev:2020dom}
V.~Dzhunushaliev and V.~Folomeev,
``Spinor field solutions in $F(B^2)$ modified Weyl gravity,''
Int.\ J.\ Mod.\ Phys.\ D {\bf 29}, no. 13, 2050094 (2020).
%2 citations counted in INSPIRE as of 15 Jan 2021

%\cite{Jizba:2020hre}
\bibitem{Jizba:2020hre}
P.~Jizba, L.~Rachwa\l{}, S.~G.~Giaccari and J.~K\v{n}ap,
``Dark side of Weyl gravity,''
Universe \textbf{6} (2020) no.8, 123;
[arXiv:2006.15596 [hep-th]].
%0 citations counted in INSPIRE as of 10 Apr 2021

\bibitem{Jizba:2019oaf}
P.~Jizba, L.~Rachwa\l{} and J.~K\v{n}ap,
``Infrared behavior of Weyl Gravity: Functional Renormalization Group approach,''
Phys. Rev. D \textbf{101} (2020) no.4, 044050;
[arXiv:1912.10271 [hep-th]].
%3 citations counted in INSPIRE as of 10 Apr 2021

\end{thebibliography}
\end{document}